\documentclass[sigconf]{acmart}
\usepackage[utf8]{inputenc}
\usepackage{subcaption}

\usepackage{xspace}

\usepackage{array}
\usepackage{balance}
\newcolumntype{L}[1]{>{\raggedright\let\newline\\\arraybackslash\hspace{0pt}}m{#1}}

\usepackage{pgf, tikz}
\usetikzlibrary{matrix,arrows,positioning,shadows,automata}
\tikzset{
    feature/.style={draw, inner sep=1.5mm, font=\small\sffamily, fill=white, drop shadow},
    opt/.style={fill=white}}
\usepackage{xcolor}

\usepackage{soul}
\usepackage{xspace}
\usepackage{cleveref}
\usepackage{footmisc}
\usepackage{multirow}
\usepackage{booktabs}
\usepackage{soul}

\newcolumntype{L}[1]{>{\raggedright\let\newline\\\arraybackslash\hspace{0pt}}m{#1}}
\newcolumntype{C}[1]{>{\centering\let\newline\\\arraybackslash\hspace{0pt}}m{#1}}
\newcolumntype{R}[1]{>{\raggedleft\let\newline\\\arraybackslash\hspace{0pt}}m{#1}}

\copyrightyear{2021} 
\acmYear{2021} 
\setcopyright{acmlicensed}
\acmConference[SIGCSE '21]{Proceedings of the 52nd ACM TechnicalSymposium on Computer Science Education}{March 13--20, 2021}{Virtual Event, USA}
\acmBooktitle{Proceedings of the 52nd ACM Technical Symposium on Computer Science Education(SIGCSE '21), March 13--20, 2021, Virtual Event, USA}
\acmPrice{15.00}
\acmDOI{10.1145/3408877.3432534}
\acmISBN{978-1-4503-8062-1/21/03}

\newcommand\taskTracker{TaskTracker}
\newcommand\tasktrackerTool{\taskTracker-tool\xspace}
\newcommand\tasktracker{\taskTracker\xspace-plugin\xspace}
\newcommand\tasktrackerServer{\taskTracker-server\xspace}
\newcommand\activityTracker{Activity Tracker\xspace}
\newcommand\dataPostprocessingTool{Data post-processing tool\xspace}

\newcommand\pluginLink{https://github.com/JetBrains-Research/task-tracker-plugin}
\newcommand\pluginWikiLink{https://github.com/JetBrains-Research/task-tracker-plugin/wiki}

\newcommand\serverLink{https://github.com/JetBrains-Research/task-tracker-server}
\newcommand\serverWikiLink{https://github.com/JetBrains-Research/task-tracker-server/wiki}
\newcommand\activityTrackerLink{https://github.com/dkandalov/activity-tracker}
\newcommand\postprocessLink{https://github.com/JetBrains-Research/task-tracker-post-processing}
\newcommand\dataLink{https://zenodo.org/record/4296613\#.X8P6jRMzZb8}

\title{\tasktrackerTool: a Toolkit for Tracking of Code Snapshots \\ and Activity Data During Solution of Programming Tasks}

\author{Elena Lyulina}
\affiliation{
  \institution{JetBrains Research}
  \country{Russia}
}
\email{elena.lyulina@jetbrains.com}

\author{Anastasiia Birillo}
\affiliation{
  \institution{JetBrains Research}
  \country{Russia}
}
\email{anastasia.birillo@jetbrains.com}

\author{Vladimir Kovalenko}
\affiliation{
  \institution{JetBrains Research}
  \country{Netherlands}
}
\email{vladimir.kovalenko@jetbrains.com}

\author{Timofey Bryksin}
\affiliation{
  \institution{JetBrains Research}
  \institution{Saint Petersburg State University}
  \country{Russia}
}
\email{timofey.bryksin@jetbrains.com}

\begin{abstract}
    The process of writing code and use of features in an integrated development environment (IDE) is a fruitful source of data in computing education research. 
    Existing studies use records of students' actions in the IDE, consecutive code snapshots, compilation events, and others, to gain deep insight into the process of student programming. 
    
    In this paper, we present a set of tools for collecting and processing data of student activity during problem-solving. 
    The first tool is a plugin for IntelliJ-based IDEs (PyCharm, IntelliJ IDEA, CLion).
    By capturing snapshots of code and IDE interaction data, it allows to analyze the process of writing code in different languages --- Python, Java, Kotlin, and C++. 
    The second tool is designed for the post-processing of data collected by the plugin and is capable of basic analysis and visualization.
    To validate and showcase the toolkit, we present a dataset collected by our tools. It consists of records of activity and IDE interaction events during solution of programming tasks by 148 participants of different ages and levels of programming experience. We propose several directions for further exploration of the dataset.
\end{abstract}



\begin{document}
\fancyhead{}

\maketitle

\section{Introduction}\label{section:introduction}
    With an ever-increasing presence of software in our lives, programming education also becomes more and more popular~\cite{mcgettrick2005grand, robins2003learning, yang2015two}. 
However, teaching programming is challenging. 
Programming courses are usually based on tasks and projects~\cite{ambrose2010learning, gratchev2018introducing} that students should complete by themselves. 
Many students may enroll for a particular programming course at once~\cite{danielsiek2016stay}. As a result, it is not always possible for teachers to pay enough attention to each individual student to have a detailed understanding of their progress.
The process of programming can be a valuable source of insight into the learning process: for example, typical ``novice errors'' may indicate particular gaps in students' understanding~\cite{altadmri201537, robins2003learning, konecki2014problems}. 
These challenges make tracking and analysis of students' coding behavior a promising technique for educational research.

A number of studies are focused on analysis of students' general behavior~\cite{jadud2005first, vihavainen2014novices}, their code patterns ~\cite{blikstein2014programming, bulmer2018visualizing}, and errors~\cite{altadmri201537, mccall2014meaningful, tirronen2015understanding}. 
Other studies aim to facilitate the process of teaching by observing the learning progress within groups in computer science (CS) courses~\cite{yan2019pensieve, blikstein2011using}.
To do so, one needs data containing interaction between a student and their programming environment. 
Such data may include sequential code snapshots, compilation events, or actions performed in the integrated development environment (IDE).

Functionality of existing data collection tools~\cite{brown2014blackbox, shah2003web, norris2008clockit, spacco2006experiences} varies depending on the purpose of their use, but they still have a lot in common. 
Such tools are usually implemented as plugins for IDEs, which facilitate the installation of the tools and allow to preserve a natural programming environment.
These tools tend to have a client-server architecture. This allows to automatically receive and store gathered data, which typically consists of user's interactions with the IDE and snapshots of their code.
However, existing tools have several restrictions. They are often tailored to support a single programming language~\cite{norris2008clockit, brown2014blackbox}, do not track the context of students' actions~\cite{norris2008clockit, spacco2006experiences, kazerouni2017deveventtracker}, or do not provide detailed information about tasks~\cite{norris2008clockit, brown2014blackbox}.
Some of the tools offer many other features besides data collection, which makes the data gathering process difficult and confusing for both students and researchers~\cite{spacco2006experiences, shah2003web}. 

Our intention was to create a tool capable of collecting all consecutive code snapshots and user actions during the \textit{solution of a programming task} in various programming languages.
Such a tool could be useful to make the process of teaching more effective by providing additional information about the process of programming. 
Detailed insights in the task solving process could help the teacher to improve their course and understand which topics and assignments may be more difficult for students. 
In addition, the tool could be used by researchers to gather data beyond the classroom.
It is important that collected activity data, such as IDE interaction events, is linked to concrete tasks so that the context of actions is available in further analysis.
Finally, the tool should be as flexible and easy to customize as possible.
In particular, data gathering should not be limited to one IDE or programming language. 

This work presents the following contributions:
\begin{itemize}
    \item \textit{\tasktrackerTool}, 
    consisting of \textit{\tasktracker}\footnote{\textit{\tasktracker}: \url{\pluginLink}} --- a plugin for gathering solution activity data in IntelliJ-based IDEs, a server\footnote{\textit{\tasktrackerServer}: \url{\serverLink}} for remote setup and gathering data, and the \textit{\dataPostprocessingTool}\footnote{\label{postprocessNote}\textit{\dataPostprocessingTool}: \url{\postprocessLink}} to perform basic analysis of the collected data of students' behaviour.
    \item A public dataset\footnote{\label{dataNote}Dataset: \url{\dataLink}} of problem-solving activity events of 148 students, retrieved and analyzed with \textit{\tasktrackerTool}, along with several suggestions for research directions where it can be of use. 
\end{itemize}

The  remainder  of  the  paper  is  organized  as  follows.  In ~\Cref{section:background}  we  discuss  related  work  and  motivate the development of \textit{\tasktrackerTool}.  
~\Cref{section:tasktracker_tool} describes the proposed tools for data gathering, post-processing, and analysis. 
~\Cref{section:dataset} describes the dataset and highlights the points of interest in a sample of data produced by \textit{\tasktrackerTool}. 
In ~\Cref{section:conclusion} we draw conclusions, discuss the encountered difficulties, and outline future work.

\section{Background}\label{section:background}
    Analysis of students' behavior in the IDE during problem-solving is an established technique in computing education research~\cite{ihantola2015educational}. Some studies rely on existing datasets for analysis~\cite{altadmri201537, mccall2014meaningful},
other researchers gather their own datasets~\cite{jadud2005first, vihavainen2014novices}.
Such datasets usually include information relevant to the programming process, such as code snapshots, interactions of users with the IDE, and various demographic data~\cite{hundhausen2017ide, ihantola2015educational}. 
Collecting such data is a tedious process that involves substantial technical work.
In this section we overview prior studies that collect such datasets with various tools and briefly describe their approaches to data collection.

Norris C. et al.~\cite{norris2008clockit} developed the \textit{Clockit} plugin for BlueJ IDE~\cite{bluej}, augmented with a web-based data visualizer, to compare behavior of novice programmers to more experienced ones visually.
Their plugin is designed to capture and log various events in the IDE, such as compilation, running the code, or changes in file size.
The collected log files are later sent to a server.
Marmoset~\cite{spacco2006experiences} is a plugin for Eclipse IDE~\cite{eclipse} that was developed as an automated testing system to grade submissions of solutions to programming tasks.
Its notable feature is automatic synchronization of contents with the version control system each time a student saves their project.
This approach enables the collection of sequential snapshots of code for further analysis.

DevEventTracker~\cite{kazerouni2017deveventtracker} is another Eclipse plugin that collects all code snapshots during the coding process.
The plugin works with the Web-CAT~\cite{shah2003web} system, which allows to analyze the data gathered in the classroom. 
The system provides features such as scoring of solutions, providing feedback to students, and tracking their progress.
Blackbox~\cite{brown2014blackbox} is a large-scale data collection project for BlueJ IDE. 
The data gathered by Blackbox includes code changes and user actions in the IDE. 
After a user consents to send their data, all changes that occur through the user's interaction with the IDE are sent to the Blackbox server. 
The authors collected a large open dataset, which is available to the community.

Existing tools provide wide opportunities for data collection and analysis.
They can be used to support students through the whole learning process or to collect comprehensive datasets with code and IDE interactions of students all around the world. 
However, existing tools also have some limitations. 

Data gathering tools that are designed to assist in teaching often include additional features, like code review or chats.
While providing additional value, such features make the tools less flexible and harder to integrate into an already established teaching workflow.
Being the plugins for IDEs such as BlueJ or Eclipse, existing tools are suitable only for a particular programming language or a relatively narrow audience~\cite{eclipse2017popularity, eclipse2020popularity}. 
Since IDEs for beginners differ from professional IDEs~\cite{hagan2000teaching, kolling2008using}, this may result in a limited scope of use for the gathered data. 
Therefore, actions performed by users and the general flow of solutions may be different in this case.
On top of that, users accustomed to other IDEs cannot participate in data gathering or use their familiar tools while learning.

Data collected by existing tools is rich enough to interpret the programming process for a variety of purposes, since it contains both user interactions with the IDE and snapshots of code.
However, sometimes code snapshots are too sparse to meaningfully restore the solution's timeline~\cite{kazerouni2017deveventtracker, spacco2006experiences, brown2014blackbox}.
Moreover, information about the task that the student is currently working on is not always complete~\cite{brown2018blackbox}. 
Such information can be essential to analyze and facilitate the process of implementation of the solution, which causes most of the difficulties for students~\cite{lahtinen2005study} and is therefore interesting to study deeply.

In our work, we strive to overcome the restrictions of existing tools, such as low flexibility, narrow range of supported IDEs, and sparsity of data. 
We present a tool called \textit{\tasktrackerTool}. 
It is designed as a plugin for IntelliJ-based IDEs~\cite{intellij} and therefore supports working with various programming languages (Python, Java, Kotlin, C++). In addition, support for any other language integrated with these IDEs can be easily added to our toolkit.

The main purpose of \textit{\tasktrackerTool} is to collect all code changes and IDE actions that happen during the process of solving programming assignments.
This data can either be used in the classroom or collected for further study in a research environment.
The user interface (UI) of the plugin allows the user to choose a task to solve. 
This feature is useful both for data gathering and for use in class: for example, the tool can be used during a test with predefined tasks to track students' individual solution patterns.
The data collected by \textit{\tasktrackerTool} is grouped per task. In contrast to existing work, this allows exploring the data while keeping track of the exact context where it was produced.
The client-server architecture allows the plugin to be used as a data gathering tool. 
Finally, the distribution package of \textit{\tasktrackerTool} includes utilities for post-processing and visualization of data, providing basic analysis capabilities and suitable for use as a base for more advanced analysis.

\section{\tasktrackerTool}\label{section:tasktracker_tool}
    Our intention was to create a tool to collect fine-grained data of how students solve their programming assignments and analyze their learning progress. 
An important requirement for such a tool, imposed by the limitations of existing approaches, is flexibility and ease of tailoring it to a particular environment. 
The tool should both extend the (currently narrow) range of IDEs that have similar plugins available and be capable of collecting snapshots of the solution's code along with information about the task.
To broaden potential use cases, the tool should not be overloaded with extra features that may affect students' behavior or require sufficient changes in teaching practices. 

We have developed a tool called \textit{\tasktrackerTool} that allows gathering sequences of code snapshots and user-to-IDE interactions during the process of solution. 
\textit{\tasktrackerTool} includes:
\begin{enumerate}
    \item \textit{\tasktracker} --- an IDE plugin to track the process of solving tasks within the IDE;
    \item \textit{\tasktrackerServer} --- a server to gather data remotely, collect solutions of multiple students in one place, and customize the  (UI) of the plugin;
    \item \textit{\dataPostprocessingTool} --- a set of utilities for basic processing of data collected by \textit{\tasktracker}.
\end{enumerate}

\begin{figure*}
    \includegraphics[width=.8\linewidth]{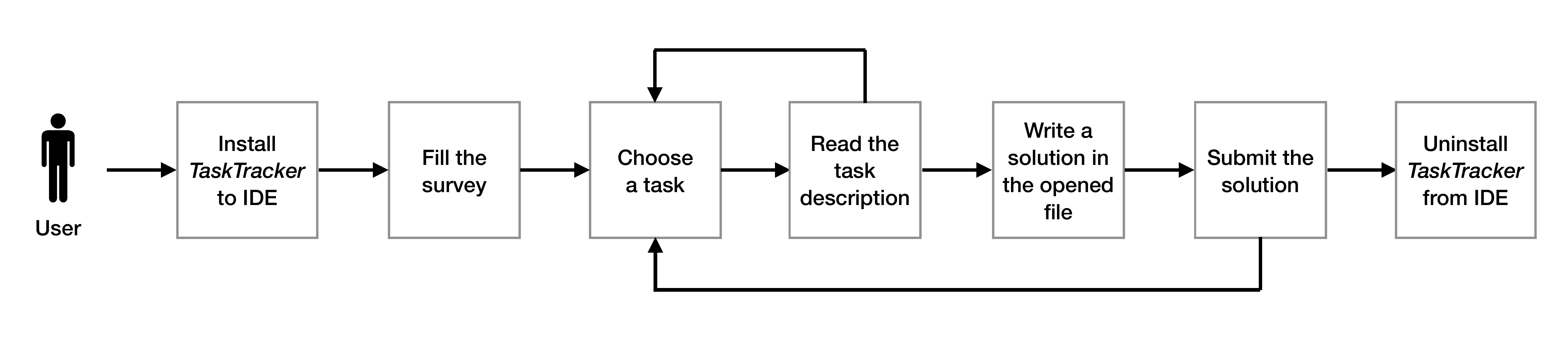}
    \caption{\textit{\tasktracker} student workflow}
    \label{fig:tasktracker_tool:tasktracker:user_workflow}
\end{figure*}

The first part of \textit{\tasktrackerTool} is \textit{\tasktracker} for IDEs based on the IntelliJ Platform~\cite{intellij}, such as PyCharm, CLion, IntelliJ IDEA, and others. 
The plugin collects all code changes during the solution process. 
In addition, it works in conjunction with the \textit{\activityTracker}~\cite{activitytracker2020tool} plugin that captures students' interaction with the IDE such as copy-paste, run-debug, and other actions. 
~\Cref{fig:tasktracker_tool:tasktracker:user_workflow} presents the workflow of a student during the plugin use. 
First of all, the plugin has to be installed into the IDE. 
The work starts with filling a survey, including gender, age, country, and programming experience. This information is then stored locally and does not need to be filled again when switching to the next task or starting the new session. Next, the student selects a task to solve from the list of tasks defined in the plugin's configuration.
Each task has its own description, which the student should read before starting to solve it. 
The plugin automatically creates a draft file for each solution.

\textit{\tasktrackerTool} has a client-server architecture. The use of a server allows to configure the plugin remotely --- for example, by updating the list of tasks --- and to collect data in a centralized manner. 
In addition, the server reduces the workload on the user's computer.
The only requirement of the plugin is the internet connection. 

The last part of \textit{\tasktrackerTool} is \textit{\dataPostprocessingTool}, a set of utilities for the analysis of the collected data. 
The task of \textit{\dataPostprocessingTool} is to prepare the gathered data for further analysis by adapting them to specific problems.
For example, it can filter out unnecessary steps in the solution flow captured by \textit{\tasktracker} or merge consecutive IDE actions and code snapshots. 
It also assists in the exploration of the gathered data through visualization of aggregated information, such as distributions of different measures of participants, the progress of individual solutions, or unusual patterns in IDE interaction.
    \subsection{\tasktracker}
        The goal of \textit{\tasktracker} is to collect all consecutive code changes and IDE actions during the solution of a programming task. 

\noindent\textbf{Collected data.}
Every time the solution is updated, even by one character, the plugin writes a snapshot of its code to the corresponding \textit{.csv} file. 
High-frequency snapshots ensure that every detail of the solution process is recorded. 
To respect students' privacy, the plugin only collects data related to the solution. To achieve this, the plugin automatically creates a dedicated file for each task where the user should write their solution, and discards changes in other files.

Interactions between the student and the IDE are tracked by an additional plugin, \textit{\activityTracker}, and are also stored in a \textit{.csv }file. The format of logs and a complete list of tracked events can be found on the project's page~\cite{activitytracker2020tool}. All the data is sent anonymously (no user names, paths or data unrelated to tasks are sent) with student's permission as soon as they mark their solution as done and submit it. 

\noindent\textbf{User interface.}
Besides data collection, the plugin provides a convenient interface to ease the routine of problem-solving for both students and teachers. 
Its UI is designed to describe every task right within the IDE so that the student is not distracted from the solution process if they have to consult with a description.
In addition, tasks can be remotely customized by the teacher for each new session.

    \subsection{\tasktrackerServer}
        \textit{\tasktrackerServer} is used in conjunction with \textit{\tasktracker} for data gathering. 
The server could be launched locally or remotely, depending on the desired setup. 
A dedicated server allows the plugin workflow to be flexible, automated, and, if needed, remote. 
The two primary functions of the server are sending data to \textit{\tasktracker} to initialize its UI and receiving data with user solutions and IDE actions. 
Remote UI configuration allows to customize the lists of tasks and language of the interface.
Remote data collection greatly facilitates the process of data gathering. The setup is as easy as cloning the server's repository, deploying it remotely or locally, and setting its URL in the plugin configuration.

\noindent\textbf{Server configuration.}
The first feature of the server is remote customization of \textit{\tasktracker}'s UI. 
The server stores a list of supported UI languages along with translations of UI texts.
Therefore, to add a new language, one should add it to the list of languages and extend existing translations.
Customization of tasks requires similar actions to the tasks list, which is also stored on the server. 
A detailed description of these and implemented data models can be found in the server repository.\footnote{\textit{\tasktrackerServer} documentation: \url{\serverWikiLink}}

\noindent\textbf{Data storage.}
The server provides another major feature --- receiving and storing files from \textit{\tasktracker}. 
The files generated by the plugin are uploaded into the server's database.
Each new user receives a unique identifier that does not disclose their identity.
This is necessary to be able to attribute solutions to a particular user later, as all data is gathered anonymously. 
At the end of the gathering process, the collected data is divided into folders per user and task and can be downloaded as an archive.
    \subsection{\dataPostprocessingTool}
        \textit{\dataPostprocessingTool} prepares raw data collected by \textit{\tasktracker} for further analysis. 
This data contains snapshots of code collected during the solution process and records of user interaction with the IDE.
The tool consists of two major modules.
The first is responsible for data processing, and the other handles data visualization.

\noindent\textbf{Post-processing.}
The data post-processing module is required to prepare raw data for further analysis. 
While post-processing could be implemented directly in the plugin, we chose to implement it as a separate set of tools for several reasons. 
First, we wanted to minimize the size of the plugin.
In addition, some operations, such as scoring of solutions, may take a long time. 
Keeping such operations, which do not have to be real-time, in the plugin could slow it down substantially. 
Finally, the separation of data processing allows to use different combinations of data processing submodules to produce different datasets. 
For example, one may want to create a family of datasets varying in degree of data granularity, or only keep a particular part of the data, such as IDE interactions or code snapshots. 

Currently, the data post-processing module consists of several submodules:
\begin{itemize}
    \item merging \textit{\activityTracker} and \textit{\taskTracker} files;
    \item scoring solutions;
    \item removing intermediate diffs.
\end{itemize}

Code snapshots and IDE interactions are collected by separate plugins and are saved into separate files. 
The data post-processing module contains a submodule for combining this data.

The submodule for scoring of solutions allows running the tests to calculate a correctness rating for each task. 
The correctness of a solution can be defined as the percentage of passing tests.
Similarly to \textit{\tasktracker}, the scoring submodule supports multiple languages: Java, Python, Kotlin, and C++.

The submodule for removing intermediate diffs is capable of deleting all intermediate, i.e. non-final, states in code snapshots that are collected during the implementation of a solution. 
The definition of ``final'' is configurable: for example, final snapshots may be taken after completion of every new line, or adjusted by other criteria.
Filtering out intermediate diffs enables adjustments to the level of data granularity, which in turn helps to reduce noise and adapt the collected data to further processing.

\noindent\textbf{Data visualization.}
The data visualization module is designed to find interesting and unusual patterns in the gathered data by plotting it.

Charts of distributions of participants and tasks are available for plotting. They may be crucial to determine whether the dataset is representative and complete. 

The next types of plots could provide deep insight into the solution process by recreating its flow.
The first plot (Figure~\ref{fig:dataset:actions_plot}) visualizes the sequence of actions performed by the user during the solution which makes them notable and transparent for analysis. 
The second plot (Figure~\ref{fig:dataset:score_plot}) represents the dynamics of the score which helps to identify problematic stages in the solution process.
It also allows to highlight characteristic patterns, like a drop in the score after an unsuccessful change.
Each of these plots may assist in the assessment of students’ understanding of the task and general programming concepts, as well as their familiarity with IDE features, and helps to tailor the learning process to suit each student best.
    \subsection{Use cases}
        Data collected by \textit{\tasktrackerTool} can find use in practical settings (for example, to support the learning process in programming courses) and in research environments (for example, one can use it to collect a dataset of the behavior of a group of students during their work on a particular task).
\textit{\tasktrackerTool} has detailed documentation.
In particular, to facilitate the ease of setup and use of the toolkit, it covers plugin setup and server deployment in detail.

\noindent\textbf{Programming courses.}
In the environment of a programming course, \textit{\tasktrackerTool} can suit several goals. 
First, thanks to remote configuration capabilities, it can serve as a framework for the observed problem-solving in the classroom.
In addition to that, analysis of the gathered data within the course group can help improve the course by tailoring the curriculum to adjust the difficulty of tasks or letting the teacher focus on topics that the students may find hard.
Moreover, during data gathering, students have an option to disclose their identifiers to the teacher.
This allows to build a personalized overview of the solution data for each student and provide individual insights. 
While the same result can be achieved by talking to students, conversations with each of them may take more time and the teacher's energy compared to the monitoring of the visualized solution process.
Finally, the plugin can help identify cheating. 
For example, if a large piece of code was inserted after executing the \textit{paste} command, which miraculously leads to a perfect score, the teacher may want to be more attentive. 
Such events can easily be highlighted in data collected by \textit{\tasktrackerTool}.

\noindent\textbf{Data gathering.}
Besides classroom use, \textit{\tasktrackerTool} can be used to collect data in a wide variety of research settings. 
The primary characteristic that makes it suitable for research data gathering is flexibility. 
\textit{\tasktrackerTool} has a client-server architecture, which enables remote data gathering. 
Moreover, UI text and tasks can be configured in any language, which does not limit the collection of data only to English-speaking participants. 
The demographic survey allows for collecting additional information about users.

\textit{\tasktracker} is capable of gathering incremental data on the process of problem-solving in different programming languages in a variety of IDEs.
The plugin is easy to install into the IDE. In addition, we provide detailed step-by-step guides for installing and uninstalling the plugin.\footnote{{\tasktracker guides: \url{\pluginWikiLink}}} 
Our own experience with using \textit{\tasktrackerTool} for data collection demonstrates that even users with little programming experience are able to install and set up \textit{\tasktrackerTool} easily.

\textit{\tasktrackerTool} is designed with care for users' privacy. 
The data is sent anonymously, and the tracking feature only extends to files automatically created by \textit{\tasktracker}. This ensures that users' privacy is respected, and protects their personal data from being unintentionally fingerprinted during data collection.

\section{Dataset}\label{section:dataset}
    \begin{figure*}
    \includegraphics[width=1\linewidth]{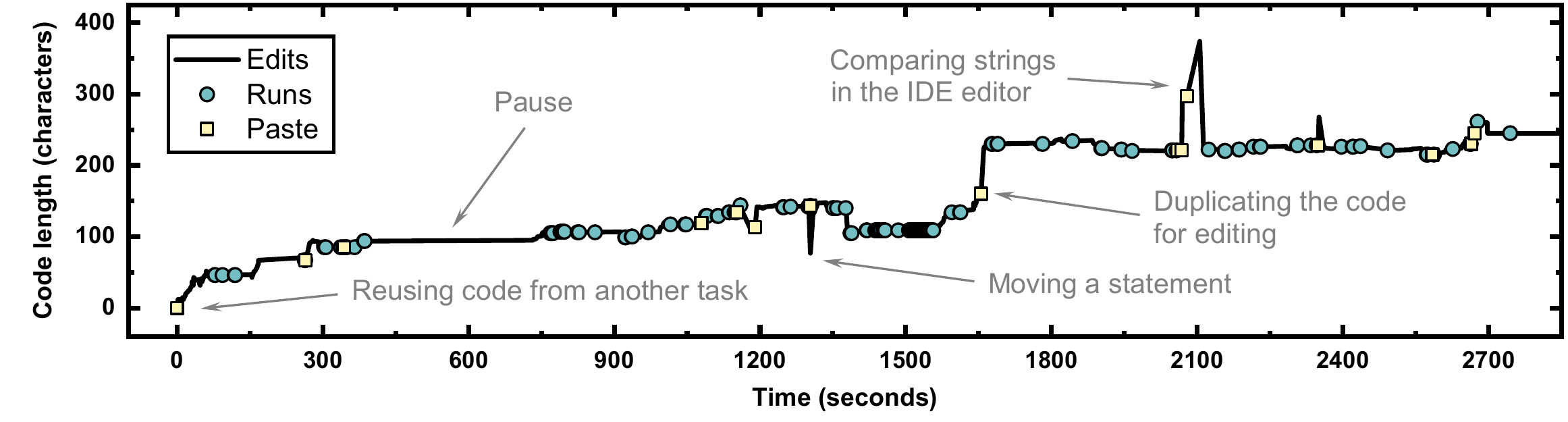}
    \caption{Actions performed in IDE during solution of a task}
    \label{fig:dataset:actions_plot}
\end{figure*}

To showcase \textit{\tasktrackerTool}, we present a dataset collected with our toolkit. While we use it in our ongoing research project, it can find use in other studies and thus may be of value to the community. We discuss some potential applications in \Cref{section:dataset:usecases}.

The dataset consists of code snapshots, IDE actions, and demographic information. 
We had 148 participants, aged 11 to 40 (mean age is 19 years), take part in the data gathering process. Their programming experience range is spanning from zero to more than 6 years of programming. 

During data gathering, solutions were accepted in one of four languages: Python, Java, Kotlin, or C++. 
However, some of the students chose not to submit tasks or solved some tasks incorrectly.
At the same time, some students solved some tasks many times in multiple languages. All submitted solutions are included in the final dataset.

For this experiment, we proposed six different programming problems of different difficulty levels as tasks. 
\Cref{tab:dataset:tasks_description} presents an overview of the problems.
Our dataset includes 474 solutions to these problems. 
\Cref{tab:dataset:solved_tasks} presents solution statistics per task and language. 

\begin{table}
\centering
     \begin{tabular}{l L{6cm} } 
         \toprule
         \multicolumn{1}{c}{\textbf{Task}} & \multicolumn{1}{c}{\textbf{Description}} \\ [0.5ex] 
         \midrule
         \textbf{Pies} & A single pie costs A dollars and B cents in the cafe. Calculate how many dollars and cents one needs to pay for N pies.  \\ 
          \midrule
         \textbf{Max 3} & Print the largest of three numbers in the input.  \\ 
          \midrule
         \textbf{Is zero} & Check if there are zeros among numbers in the input.  \\ 
          \midrule
         \textbf{Voting} & Given three numbers, each of them being 1 or 0, determine which one occurs more often: 1 or 0. Print the number that occurs more often.  \\ 
          \midrule
         \textbf{Max digit} & Given a string containing only digits, find and print the largest digit.  \\ 
          \midrule
         \textbf{Brackets} & Place opening and closing brackets into the input string like this:\newline for odd length: example → e(x(a(m)p)l)e;\newline for even length: card → c(ar)d, but not c(a()r)d.  \\ 
         \bottomrule
    \end{tabular}
    \vspace{2mm}
    \caption{Task descriptions}
    \vspace{-5mm}
    \label{tab:dataset:tasks_description}
\end{table}

\begin{table}
\centering
    \setlength{\tabcolsep}{3.4pt}
     \begin{tabular}{L{1.5cm} | C{0.5cm} C{0.5cm} | C{0.4cm} C{0.4cm} | C{0.38cm} C{0.38cm} | C{0.38cm} C{0.38cm} | C{0.49cm} C{0.49cm}} 
         \toprule
         \multirow{2}{*}{\textbf{Task}} & \multicolumn{2}{c|}{\textbf{Python}} & \multicolumn{2}{c|}{\textbf{Java}} & \multicolumn{2}{c|}{\textbf{Kotlin}} & \multicolumn{2}{c|}{\textbf{C++}} & \multicolumn{2}{c}{\textbf{All}} \\ [0.4ex]
         & \textit{S} & \textit{NS} & \textit{S} & \textit{NS} & \textit{S} & \textit{NS} & \textit{S} & \textit{NS} & \textit{S} & \textit{NS} \\
         \midrule 
         \textbf{Pies}   &     67 & 31           & 16 & 7            & 4 & 4               & 1 & 0           & \textbf{88} & \textbf{42} \\ 
         \textbf{Max 3}  &     31 & 12           & 22 & 2            & 4 & 4               & 1 & 0           & \textbf{58} & \textbf{18} \\ 
         \textbf{Is zero}&     27 & 13           & 8 & 0            & 7 & 1               & 2 & 0            & \textbf{44} & \textbf{14}\\
         \textbf{Voting} &     32 & 29           & 24 & 0            & 3 & 4              & 1 & 0           & \textbf{60} & \textbf{33}\\
         \textbf{Max digit}&   16 & 5           & 5 & 1            & 5 & 4              & 1 & 0           & \textbf{27} & \textbf{10} \\
         \textbf{Brackets} &   37 & 28           & 4 & 1            & 6 & 2              & 2 & 0           & \textbf{49} & \textbf{31} \\
         \midrule
         \textbf{All}      &   \textbf{210} & \textbf{118} & \textbf{79} & \textbf{11}   & \textbf{29} & \textbf{19}     & \textbf{8} & \textbf{0}  & \textbf{326} & \textbf{148} \\
         \bottomrule
    \end{tabular}
    \vspace{2mm}
    \caption{Number of submitted solutions: \textit{S} --- number of correct solutions; \textit{NS} --- number of incorrect solutions}
    \vspace{-7mm}
    \label{tab:dataset:solved_tasks}
\end{table}

\begin{figure}
    \includegraphics[width=1\linewidth]{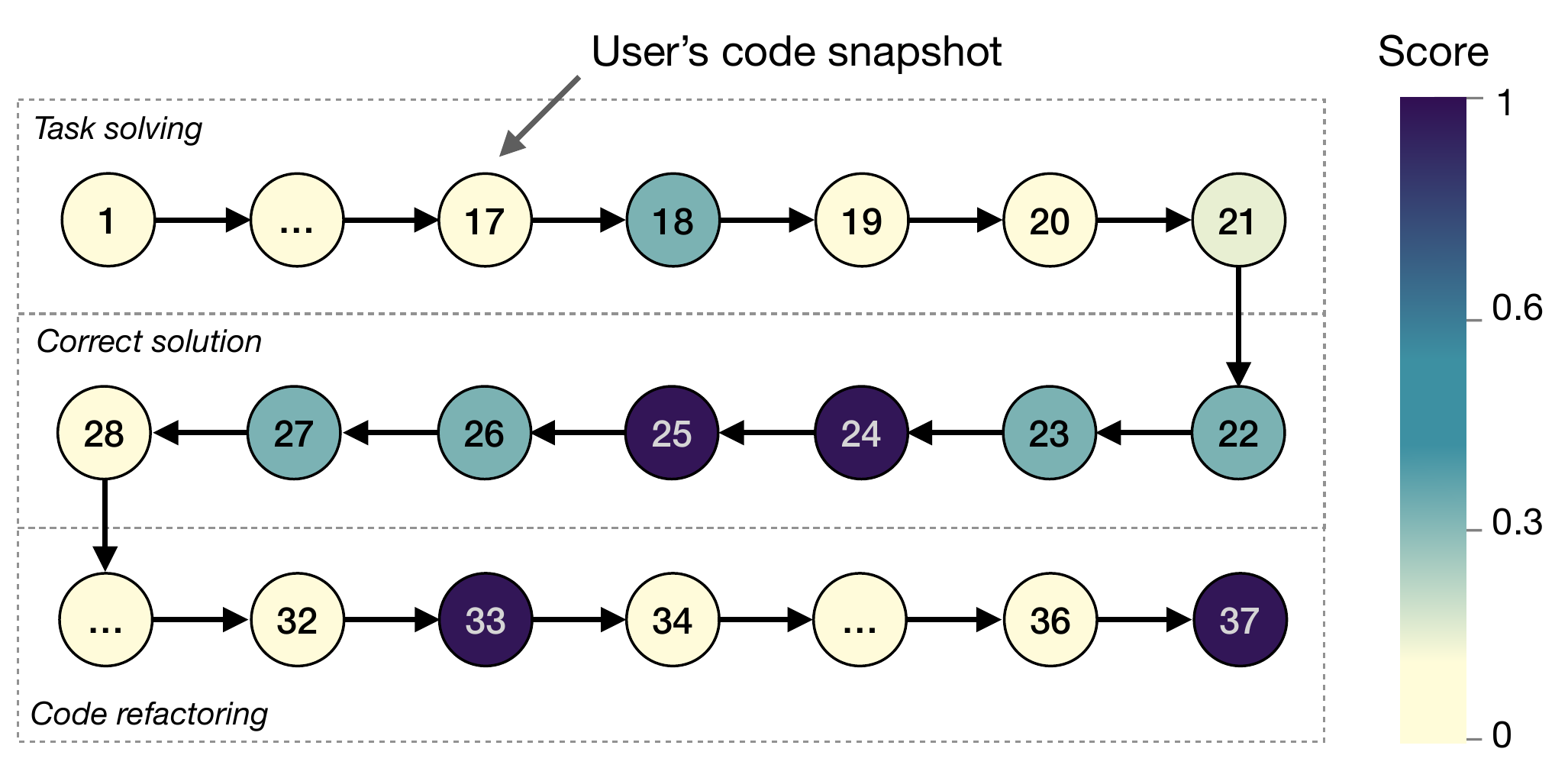}
    \caption{Changes of task score during a sample solution}
    \label{fig:dataset:score_plot}
\end{figure}
    \subsection{Use cases}\label{section:dataset:usecases}
        The gathered dataset is publicly available.
To preserve privacy, we anonymized variables and function names in the code; IDs of users are also depersonalized. 
In this subsection, we present several topics for exploration in our dataset, which other researchers may find of interest.

\noindent\textbf{Experience and feature use.}
The dataset can be used to investigate the relationship between programming experience and the use of language features or IDE actions.
Beginners may not yet be familiar with some basic features of an IDE, such as a debugger. 
This may influence their productivity during problem-solving and pose a barrier for further learning.

\noindent\textbf{Influence of age on feature use.}
It is not only children who learn to program. 
In a group of students with similar programming experience, ages may differ greatly. 
It would be interesting to see how the use of language features and patterns of actions performed in the IDE is related to age.

\noindent\textbf{Actions after solution.}
After getting the solution right, students do not always submit the task immediately: for example, they may first refactor the code to improve it. 
It could be interesting to know the most frequent changes and IDE actions that occur after achieving a perfect score.

\noindent\textbf{Common errors.}
While our tasks may look trivial to experienced programmers, even participants with impressive experience still made mistakes during the solution. 
However, common errors may be different for people with different experiences.
A deeper study of error patterns for different experience groups could be another potential application of our dataset.

\noindent\textbf{Advanced solution metrics.}
Solution snapshots in our dataset could be evaluated using different measures. 
For example, one could derive a total count of characters added or removed during the solution, or a ratio of the time spent writing code to the total time since the start of the solution.
Such metrics could highlight individual traits of students and potentially be used to personalize their learning process and environment.

\noindent\textbf{Generating personalized hints.}
Another interesting application of the dataset that we use in our ongoing research project is generation of personalized hints. Changes of code can be used to build a model of the solution process of a given task and suggest personalized hints to help students stuck in common pitfalls to make the next step towards a correct solution. Therefore, both the information about the current task and consequent code snapshots are important sources of data. To the best of our knowledge, there is no such dataset for high-level programming languages, so we collected our own.

    \subsection{Analysis}
        We analyzed the collected dataset using the \textit{\dataPostprocessingTool} to explore users' behavior and programming patterns.

\noindent\textbf{IDE actions.}
To look closely at actions performed in IDE while solving tasks, we plotted them together with the length of the current code fragment. 
~\Cref{fig:dataset:actions_plot} presents an example of such a plot for one of the solutions for the \textit{brackets} task.
Descriptions and arrows were added manually for clarification.

At the beginning of the solution, the user pasted a large piece of code which turned out to be the keyboard input code from a previous solution.
Further, the user took a break for about 6.5 minutes, an explanation of which may be the user trying to come up with a solution before starting to code. 
Another interesting detail is an abundance of code runs, which is quite unusual for most users.
This behavior might highlight their preference to have the solution code valid at all stages, even while it is still incomplete.
Also, we found that in the middle of the process the user pasted actual and expected outputs of their program for character-by-character comparison, thus using the editor as a diff tool.

\noindent\textbf{Score changes.}
Another way to analyze users' behavior is to explore the dynamics of their solution score. 
Using our post-processing tool, we removed all intermediate code changes and colored each solution state according to its score. 
The plot of this type in Figure~\ref{fig:dataset:score_plot} shows that the process of this solution can be separated into three stages: implementing an initial solution, correcting it, and refactoring. 
Analysis of the score dynamics alongside snapshots of source code allows to reason about students' intent in detail, thus providing deep insights into their approach to each individual solution.

\section{Conclusion}\label{section:conclusion}
    In this paper, we introduced \textit{\tasktrackerTool} --- a toolkit for collecting exhaustive code snapshots and user actions during the solution of programming tasks in various programming languages in IntelliJ-based IDEs. 
The toolkit enables its user to gather solution data separated per individual task, collect it on a server, process the data, and perform basic analysis over it.
Using the toolkit, we collected a dataset of solution activity by 148 participants with different experience levels.
The dataset is publicly available.
We found several interesting patterns in the dataset and suggested multiple directions for future use of the dataset and our toolkit. 

Future work on \textit{\tasktrackerTool} would involve adding support for collecting programming data in arbitrary environments to allow tracking of code in settings beyond predefined tasks while respecting privacy. 
Apart from that, we are planning to extend our dataset to more participants.
Finally, we are going to provide support for further development of \textit{\tasktrackerTool} and its adaptation to other researchers' needs, if the community considers the toolkit valuable.

\bibliographystyle{ACM-Reference-Format}
\balance
\bibliography{cite}

\end{document}